\documentclass[11pt]{article}
\usepackage{mathtools}
\usepackage{amsmath, amsthm, amsfonts,amssymb}
\usepackage{enumitem}
\usepackage{comment}
\usepackage{graphicx}
\usepackage{colortbl}
\usepackage{tikz}
\usepackage[utf8]{inputenc}
\usepackage{esint}
\usepackage{mathrsfs}
\usepackage[T1]{fontenc}
\usepackage{mathrsfs}
\usepackage{varwidth}
\usepackage{pgfplots}
\usetikzlibrary{intersections, pgfplots.fillbetween}
\usepackage{tikz}
\usepackage{tikzit}

\tikzstyle{GreenNeuron}=[fill={rgb,255: red,0; green,128; blue,128}, draw={rgb,255: red,0; green,128; blue,128}, shape=circle]
\tikzstyle{BlueNeuron}=[fill={rgb,255: red,0; green,0; blue,173}, draw=none, shape=circle]
\tikzstyle{RedNeuron}=[fill={rgb,255: red,202; green,0; blue,0}, draw=none, shape=circle]
\tikzstyle{BlackCirc}=[fill=black, draw=black, shape=circle, font={\tiny}, scale=0.4]
\tikzstyle{tinyGreen}=[fill={rgb,255: red,0; green,128; blue,128}, draw={rgb,255: red,0; green,128; blue,128}, shape=circle, scale=0.4]
\tikzstyle{tinyRed}=[fill={rgb,255: red,202; green,0; blue,0}, draw={rgb,255: red,202; green,0; blue,0}, shape=circle, scale=0.4]
\tikzstyle{RedDisc}=[fill=none, draw={rgb,255: red,191; green,0; blue,64}, shape=circle, scale=0.75]
\tikzstyle{smallRedDisk}=[fill=none, draw={rgb,255: red,191; green,0; blue,64}, shape=circle, scale=0.6]
\tikzstyle{BigRedDisk}=[fill=none, draw={rgb,255: red,191; green,0; blue,64}, shape=circle, scale=0.9]
\tikzstyle{SuperBigDisckRed}=[fill=none, draw={rgb,255: red,191; green,0; blue,64}, shape=circle, scale=1.05]
\tikzstyle{tinyBlack}=[fill=black, draw=black, shape=circle, scale=0.4]

\tikzstyle{Arrow}=[->, draw=black]
\tikzstyle{Green}=[-, draw={rgb,255: red,0; green,128; blue,128}, thick]
\tikzstyle{GreenDashed}=[-, draw={rgb,255: red,128; green,128; blue,128}, thick, dashed]
\tikzstyle{Dashed}=[-, dashed]
\tikzstyle{RedDashed}=[-, thick, dashed, draw={rgb,255: red,202; green,0; blue,0}]
\tikzstyle{BlueDashed}=[-, draw={rgb,255: red,0; green,0; blue,173}, thick, dashed]
\tikzstyle{dotRed}=[-, draw={rgb,255: red,202; green,0; blue,0}, dotted]
\tikzstyle{Thikblue}=[-, draw={rgb,255: red,0; green,0; blue,173}, thick]
\tikzstyle{ArrowRed}=[draw={rgb,255: red,191; green,0; blue,64}, ->]
\tikzstyle{doubleRedArrow}=[->, draw={rgb,255: red,191; green,0; blue,64}, double]
\tikzstyle{blueArrow}=[draw=blue, ->]
\tikzstyle{spring}=[-, spring]
\tikzstyle{LargeSpring}=[-, largeSpring]
\tikzstyle{thikRedArrow}=[->, thick, draw={rgb,255: red,191; green,0; blue,64}]
\tikzstyle{empty}=[-, draw=none]
\tikzstyle{greenArrow}=[->, draw={rgb,255: red,0; green,128; blue,128}]
\tikzstyle{dubleArrow}=[<->]
\tikzstyle{dashedArrow}=[dashed, ->]
\tikzstyle{double Green Arrow}=[<->, draw={rgb,255: red,0; green,128; blue,128}]
\tikzstyle{RedDashedFilled}=[-, draw={rgb,255: red,212; green,5; blue,9}, dashed, thick, fill={rgb,255: red,255; green,203; blue,194}, tikzit fill={rgb,255: red,255; green,203; blue,194}]
\tikzstyle{new edge style 0}=[-, draw={rgb,255: red,20; green,179; blue,52}]
\tikzstyle{new edge style 1}=[-, draw={rgb,255: red,231; green,0; blue,0}, thick=1]

\usepackage{bbm}
\usepackage{enumitem}
\usepackage{biblatex} 
\usepackage{mathtools}
\usepackage{array}
\usepackage{graphicx}
\usepackage{caption}
\usepackage{subcaption}
\usepackage{dsfont}
\usepackage{ushort}
\usepackage[labelfont=bf]{caption}
\usepackage{tikz-cd}
\usepackage[framemethod=tikz]{mdframed} 
\usepackage{tikz}
\usepackage{accents}
\usepackage{booktabs}
\usepackage{nicefrac}

\usepackage{tikz}
\usetikzlibrary{angles, quotes,calc,patterns}
\usepackage{tikz}
\usepackage{tikz-3dplot}
\usepackage{lineno} 
\usepackage{mathtools}
\numberwithin{equation}{section}
\pgfplotsset{compat = newest}
\usepackage[titletoc, toc, page]{appendix} 

\newcommand{\be}{\begin{equation}}
\newcommand{\ee}{\end{equation}}

\newcommand{\eps}{\varepsilon}

\newcommand\blfootnote[1]{%
	\begingroup
	\renewcommand\thefootnote{}\footnote{#1}%
	\addtocounter{footnote}{-1}%
	\endgroup
}

\usepackage[colorlinks=true,linkcolor=blue,citecolor=blue,urlcolor=blue,breaklinks]{hyperref}
\usepackage{hyperref}
\usepackage{url}

\def\1{\mathds{1}}

\let\eps\varepsilon

\theoremstyle{definition}

\theoremstyle{remark}

\hypersetup{pdftitle={Anisotropic}}
\hypersetup{pdfauthor={Merino-Aceituno, Moschella, Otsuka, Schmeiser, Scholz}}

\author{%
Sara Merino-Aceituno%
  \footnote{Faculty of Mathematics, University of Vienna, Oskar-Morgenstern-Platz 1,
            1090 Vienna, Austria.\\
            \texttt{sara.merino@univie.ac.at},
            \texttt{carmela.moschella@univie.ac.at},
            \texttt{christian.schmeiser@univie.ac.at}}%
\and
Carmela Moschella{{$^* $}}%
  \footnote{Vienna BioCenter PhD Program, Doctoral School of the University of Vienna
            and Medical University of Vienna, 1030 Vienna, Austria.}%
\and
Shotaro Otsuka{{$^\text{§}$}}%
  \footnote{Max Perutz Labs, Vienna BioCenter Campus, Campus-Vienna-BioCenter 1,
            1030 Vienna, Austria.\\
            \texttt{shotaro.otsuka@maxperutzlabs.ac.at}}%
\and
Christian Schmeiser{{$^*$}}%
\and
Julia Scholz{{$^{\dag \ddag}$}}%
  \footnote{Medical University of Vienna, Max Perutz Labs, Dr. Bohr Gasse 9, 1030 Vienna, Austria.\\
            \texttt{julia.scholz@meduniwien.ac.at}}%
  \footnotemark[\value{footnote}]%
}

\title{Modeling Protein Diffusion Across ER–Nuclear Envelope Junctions Reveals Efficient Transport via Simple Diffusion}

\makeindex

\begin{document}
\maketitle
	
	\vspace{-10pt}

	\begin{abstract}
		\noindent
The endoplasmic reticulum (ER) is the largest continuous membrane-bound organelle in the cell and plays a central role in the synthesis and turnover of many lipids and proteins. It connects directly to the nucleus through specialized contact points known as ER–nuclear envelope (NE) junctions. In our recent study, we found that these ER–NE junctions are both narrow and infrequent, measuring less than 20 nanometers in diameter and occurring at a frequency of approximately 0.1 per square micrometer. However, it remains unclear whether such limited and narrow connections are sufficient to support efficient transport between the ER and NE. Here, we built a mathematical model of ER-to-NE protein diffusion, incorporating ultrastructural parameters, the frequency of ER–NE junctions, and the diffusion coefficient of proteins within the ER lumen. To validate the model, we experimentally quantified the transport rate of ER luminal proteins to the NE using fluorescence recovery after photobleaching (FRAP). Our model and experimental data demonstrate that simple diffusion is sufficient to account for the rapid transport of proteins from the ER to the NE, despite the limited and narrow nature of the connecting junctions. 
Together, these findings offer mechanistic insight into how ER–NE connectivity enables rapid protein transport and lay the groundwork for future studies on ER–nucleus communication.

	\end{abstract}
	
	\tableofcontents

\section{Introduction }
The endoplasmic reticulum (ER) is a central organelle in eukaryotic cells, responsible for protein synthesis and quality control, lipid metabolism, and calcium storage. The nuclear envelope (NE), a specialized subdomain of the ER, encloses the nucleus and plays a crucial role in nucleocytoplasmic transport and gene regulation \cite{Ellenberg1997, Pawar2021}. The NE comprises inner and outer membranes, with the outer membrane being continuous with the ER \cite{Watson1955, Whaley1960, Ungricht2017, Deolal2024}. Although the exchange of membrane components and proteins between the ER and NE is vital for cellular homeostasis, the ultrastructural details of the physical connections supporting this exchange have only recently been described \cite{bragulat2024endoplasmic}.

Studies using electron microscopy have revealed that the ER--NE junctions are spatially restricted and structurally narrow, typically less than 20\,nm in diameter in plant and mammalian cells \cite{Craig1988, Staehelin1997, bragulat2024endoplasmic}. These junctions are also extremely sparse, with a spatial density of approximately 0.1 junctions/$\mu$m$^2$ \cite{bragulat2024endoplasmic}, which is significantly lower than the density of nuclear pores (5--10 pores/$\mu$m$^2$) \cite{Otsuka2016}. These structural constraints raise questions about how efficient transport can be achieved through such geometrically restricted junctions. Specifically, it remains unclear whether these narrow and rare junctions are capable of supporting sufficient luminal protein transport by passive diffusion alone or whether active transport processes might be required.

Previous research on ER luminal transport has largely focused on the mobility within the ER and how it is influenced by the ER architecture \cite{Dayel1999, Holcman2018, Htet2024, Crapart2024, Konno2024}. However, the dynamics of luminal transport between the ER and NE remain poorly characterized. In this study, we address this gap by developing a model of protein diffusion from the ER to the NE. The model incorporates known geometric constraints imposed by the architecture of ER–NE junctions to predict the transport rate of proteins into the NE. To validate the model, we used fluorescence recovery after photobleaching (FRAP) to monitor the dynamics of ER luminal reporters of different sizes in live HeLa cells. By comparing experimental recovery curves to model simulations, we found that passive diffusion alone can account for the observed transport dynamics, even through rare and narrow ER--NE junctions.

\section{Materials and methods}\label{sec:mat-meth}

\subsection{Cell culture}\label{sec:cell_culture} 
A plasmid encoding moxGFP-KDEL was a gift from Erik Snapp (Addgene plasmid \#68072; \url{http://n2t.net/addgene:68072}; RRID:Addgene\_68072; \cite{Costantini2015}). To generate NusA-moxGFPx2-KDEL, an HA-tagged NusA fragment was first amplified by PCR from the plasmid encoding 4xHA-NusA (a kind gift from J. Ellenberg lab, European Molecular Biology Laboratory) using the primers 5$'$-ACGACCGGTA-

\noindent GCCACCATGGGATATCCCTACG-3$'$ (forward) and 5$'$-TCGACCGGTAAGCTTGCCGCTTCGTCA-

\noindent CCGAACCAG-3$'$ (reverse). The amplified fragment was then inserted into the moxGFP-KDEL plasmid using AgeI (New England Biolabs). Subsequently, a second moxGFP fragment was amplified from moxGFP-KDEL by PCR using the primers 5$'$-AGGAAGCTTGGATCCAGTGTCCAAGGGCGAGG-

\noindent AGCTG-3$'$ (forward) and 5$'$-TAGAAGCTTGCCTTGTACAGCTCGTCCATGCCGT-3$'$ (reverse), and inserted between the HA-NusA and moxGFP-KDEL sequence using HindIII (New England Biolabs).

\subsection{Live cell imaging and photobleaching }
moxGFP-KDEL was introduced by transient transfection with Lipofectamine 3000 (Thermo Fisher Scientific, L3000001). NusA-moxGFP$\times$2-KDEL was delivered either by transient transfection or via a lentiviral vector system using pMD2.G and pSPAX (gift from R. Foisner lab, Max Perutz Labs). 
At least 1 hour prior to imaging, cells were washed with 1X PBS (Sigma-Aldrich, D8537-500ML) and the medium was exchanged to imaging media (DMEM without Riboflavin and Phenol Red, Thermo Fisher Scientific, Gibco 041-96205M), containing 10\% FBS, 1\% penicillin/streptomycin, and 50nM SiR-DNA (Spirochrome, SC007). Image acquisition and photobleaching experiments were performed by spinning disc microscopy (Yokogawa CSU-X1-A1 Nipkow) with a Plan-Apochromat 63x/1.4 NA oil immersion objective (Carl Zeiss), maintained at 37°C with 5\% CO2 in a microscope-body-enclosing incubator. For the control experiment that estimates the bleaching depth, cells were fixed with 4\% paraformaldehyde (PFA) in phosphate-buffered saline (PBS) at 37°C for 30 min, followed by replacement with fresh PFA for an additional 30 minutes. After multiple washes in PBS, cells were incubated with imaging media and imaged in 3D using spinning disc microscopy under the following conditions: 69 optical sections, z-stacks every 250 nm, and xy pixel size of 212 nm.

To measure diffusion coefficients, a small region of the ER was photobleached using a 488 nm laser, and images were acquired every 0.25 seconds to monitor fluorescence recovery. For moxGFP-KDEL a square region (4.23 x 4.23 $\mu$m ) was photobleached, while for NusA-moxGFPx2-KDEL, rectangular regions (1.48 x 5.29 $\mu$m or 7.4 x 7.4 $\mu$m ) were used.  For ER-to-NE transport measurements, the entire nucleus was photobleached, and recovery of fluorescence was tracked every 1 second. SiR-DNA signal was simultaneously recorded using a 640 nm laser to define the nuclear periphery for subsequent image analysis.

\subsection{Quantification of diffusion coefficient and transport rate constant} 
We used the ImageJ plugin simFRAP \cite{Blumenthal2015} to measure the diffusion coefficient within the ER. Briefly, the polygon tool was used to select the whole ER (excluding the nucleus) of the bleached cell and of a reference (unbleached) cell. A rectangle was used to select the bleached area (moxGFP-KDEL: 3.60 $\times$ 3.60 $\mu$m, NusA-moxGFPx2-KDEL: 1.27 $\times$ 4.87 $\mu$m).

To quantify ER-to-NE transport, fluorescence intensity was measured at each frame of the time-lapse images at the NE. An NE mask with a width of 2 pixels (corresponding to 423 nm) was generated from the SiR-DNA channel. To do this, the channel was first median-filtered (radius = 3 pixels), then thresholded, dilated by 2 pixels, and finally subtracted from the original mask. The intensity values were corrected for photobleaching by using non-bleached cells as a reference after background subtraction. The intensity at the first post-bleach time point was normalized to zero, and the upper plateau was normalized to the fitted value obtained by applying a single exponential decay model (using a one-phase association in GraphPad Prism version 9.02 for Windows, GraphPad Software, www.graphpad.com). The normalized intensity curves were then aligned to the photobleaching time point and averaged across cells. All image analyses were performed using Fiji \cite{schindelin2012fiji}. 

\subsection{Shape of ER-NE junctions}
3D ultrastructure of ER-NE junctions were visualized using electron tomography and the 3D meshes were generated in a previous study \cite{bragulat2024endoplasmic}. Cross-sections were taken at regular intervals along the axis of each 3D junction meshes. The radius was determined from the cross-section with the smallest surface area. To obtain the average junction shape, cross-sectional areas were measured along the junction axis for each mesh and aligned based on their minimal area. The averaged cross-sectional area was then calculated from –2.5 nm to 10 nm along the junction axis. Finally, the area was converted to radius and plotted.

\subsection{The mathematical model}
The dynamics of the unbleached fluorescent reporters will be described, concentrating on the junctions between the ER and the NE. The densities within the ER and the NE are assumed to be approximately homogeneous and their values at time $t$ are denoted by $\rho_{ER}(t)$ and $\rho_{NE}(t)$. For simplicity, the junctions are assumed to have identical shapes and equal dynamics, such that it is sufficient to describe one of them. As a further simplification, the protein density in the junctions is assumed to be constant along cross sections. This can be justified for junctions with large aspect ratios (see the appendix), but it is also used here for junctions whose length and diameter are of the same order of magnitude. Note, however, that the junction length $L$ is not well defined, as there is no clear criterion for determining where the junction ends. The diffusion through a junction can then be described by a one-dimensional model, where the protein density $\rho_J(z,t)$ in the junction depends only on the longitudinal variable $z$, varying between $z = 0$ and $z = L$, corresponding to the connections with the NE and, respectively, the ER. The diffusive flux through the junction cross section at position $z$ and time $t$ is then given by the Fourier law $-D A(z) \frac{\partial\rho_J}{\partial z}(z,t)$ with the diffusivity $D$ and the cross section area $A(z)$. Therefore the one-dimensional density $A\rho_J$ solves the diffusion equation
\be\label{rhoJ}
   A(z)\frac{\partial \rho_J}{\partial t}(z,t) = \frac{\partial}{\partial z}\left( D A(z) \frac{\partial\rho_J}{\partial z}(z,t)\right) \,.
\ee
The amount of protein in the ER changes only by the flux in and out of the 
junctions, and therefore
\be\label{rhoER}
   V_{ER} \frac{d\rho_{ER}}{dt}(t) = -k D A(L) \frac{\partial\rho_J}{\partial z}(L,t) \,,
\ee
with the total ER volume $V_{ER}$ and with the number $k$ of junctions. Analogously, for the NE:
\be\label{rhoNE}
   V_{NE} \frac{d\rho_{NE}}{dt}(t) = k D A(0) \frac{\partial\rho_J}{\partial z}(0,t) \,.
\ee
The signs in \eqref{rhoER} and \eqref{rhoNE} are a consequence of the $z$-direction pointing into the ER and out of the NE.

The equations \eqref{rhoJ}--\eqref{rhoNE} are a complete model for the densities $\rho_J$, $\rho_{ER}$, and $\rho_{NE}$, if complemented by initial conditions and the natural boundary conditions at the ends of the junction:
\be\label{BC}
   \rho_J(0,t) = \rho_{NE}(t) \,,\qquad \rho_J(L,t) = \rho_{ER}(t) \,.
\ee
A further simplification results from the fact that the total volume of the ER
is much bigger than the volume of the NE, and this in turn is much bigger than the total volume of the junctions. This has the consequence that the differential equations \eqref{rhoJ}--\eqref{rhoNE} act on 3 strongly different time scales. The dynamics in the junctions is much faster than the dynamics in the NE, which in turn is much faster than the dynamics in the ER. Since the experimental observations are concerned with the NE, we concentrate on the intermediate time scale and approximate the fast dynamics in the junctions by a quasistationary model, replacing \eqref{rhoJ} by
\be\label{rhoJx}
   0 = \frac{\partial}{\partial z}\left( D A(z) \frac{\partial\rho_J}{\partial z}(z,t)\right) \,,
\ee
and the slow dynamics in the ER by 
\be\label{rhoERx}
   \frac{d\rho_{ER}}{dt}(t) = 0 \,,
\ee
instead of \eqref{rhoER}.

The model contains all the cross section areas of the junction between the connections to the ER and to the NE. Only an average value will be needed, which is computed by the inverse mean value:
   \begin{equation}\label{Astar}
       \frac{1}{A^*} = \frac{1}{L}\int_0^L \frac{dz}{A(z)} \,.
    \end{equation}
The average radius $R^*$ then satisfies the formula $A^* = R^*\pi(R^*+L\tan\alpha)$. The values for $L$ and $R^*$ in the table are rough estimates, obtained from visual inspection of EM representations of several junctions. For $R^*$, we use an effective radius defined as the difference between the radius of the junction and that of the reporter, since the reporter’s size is of the same order of magnitude as the junction radius. Specifically, we will compute $R^*= R-r_s$ for the small reporter and $R^*=R-r_l$ for the large reporter following \textbf{Table \ref{tab:table1}}.
For more details on the model derivation we refer to the appendix, in particular for the derivation of the one-dimensional model \eqref{rhoJ} and the passage from \eqref{rhoJ}--\eqref{BC} to \eqref{rhoNE}--\eqref{rhoERx} by formal asymptotic methods.

\section{Results}
    \subsection{Live-cell reporters for quantifying luminal protein flux}
    We used two reporters of different sizes to quantify protein flux between the ER and NE lumen (\textbf{Figure 1A}). The first is moxGFP, a monomeric GFP variant optimized for oxidizing environments such as the ER \cite{Costantini2015}, fused with the ER retention signal KDEL. This smaller reporter has a molecular weight of approximately 30 kDa. The second, referred to as the larger reporter, consists of two tandem moxGFP fused to the inert, bulky protein NusA and KDEL, with a total molecular weight of about 120 kDa. When expressed in HeLa cells, both reporters showed even localization throughout the ER and NE (\textbf{Figure 1B}). Photobleaching of small regions within the ER showed that the smaller reporter diffuses rapidly throughout the ER lumen (\textbf{Figure 1B}). Quantification of fluorescence recovery closely matched simulated diffusion kinetics \textbf{(Figure 1C)}. The apparent diffusion coefficient was 3.3$\pm$1.3 $\mu m ^2$/sec (mean $\pm$ s.d., n = 7 cells), which is consistent with reported values for GFP within the ER \cite{Dayel1999}. In contrast, the larger reporter had a significantly lower diffusion coefficient of 0.52$\pm$0.44 $\mu$m²/s (n = 17 cells) (\textbf{Figure 1C}), likely due to its elongated shape and increased molecular size. These results show that both reporters serve as effective tools for studying how protein size influences diffusion dynamics within the ER and NE lumen in live cells.

    \subsection{Quantifying ER-to-NE protein flux using photobleaching}
    The ER and NE lumen are connected through narrow membranous junctions (\textbf{Figure 1A}). To measure protein flux from the ER to the NE, we photobleached the reporters specifically within the NE, and then monitored how unbleached reporters from the ER accumulate in the NE over time. To ensure complete photobleaching of the entire NE in 3D, we repeated the experiment in fixed cells where no diffusion can occur, and confirmed that our bleaching conditions were effective (\textbf{Figure 1D}). In live cells, both the smaller and larger reporters rapidly diffused into the NE following bleaching. The smaller reporter reached a steady-state level within 20 seconds, while the larger reporter did so within 2 minutes (\textbf{Figure 1E,F}). Given that the junctions connecting the ER to the NE are narrow and infrequent \cite{bragulat2024endoplasmic}, we next investigated whether the observed rapid flux could be explained by passive diffusion alone or whether active transport mechanisms were involved.

\subsection{Prediction of the transport rate by the mathematical model}

The model \eqref{rhoNE}--\eqref{rhoERx} is simple enough to be solved explicitly.
Equation \eqref{rhoJx} implies that the diffusive flux in the junction is independent from the position $z$:
$$
  j_J(t) = -D A(z) \frac{\partial\rho_J}{\partial z}(z,t) \,.
$$
Division by $DA$, integration with respect to $z$, and the boundary conditions \eqref{BC} imply
$$
  j_J(t) = -DA^* \,\frac{\rho_{ER}(t) - \rho_{NE}(t)}{L} \,,
$$
with the average cross section area \eqref{Astar}. From \eqref{rhoERx} we have
$$
  \rho_{ER}(t) = \rho_{ER,0} \,,
$$
the constant value of the density in the ER. With these results, \eqref{rhoNE} can be written as
\be\label{kappa}
   \frac{d\rho_{NE}}{dt}(t) = \kappa \bigl(\rho_{ER,0} - \rho_{NE}(t)\bigr) \,,\qquad\text{with } \kappa = \frac{kD A^*}{V_{NE}L} \,.
\ee
If $t=0$ denotes the end of the bleaching process, implying $\rho_{NE}(0)=0$, then the solution of this differential equation is given by
$$
\rho_{NE}(t) = \rho_{ER,0}\left(1-e^{-\kappa t}\right) \,,
$$
showing that $\kappa$ is the desired prediction of the transport rate constant.

\subsection{Comparison Between Experimental Observations and Model Predictions}
We formulated a mathematical model that describes the flux of proteins between the ER and the NE, using a differential equation and incorporating structural constraints at the ER–NE junctions (\textbf{Figure 2A}, details in Methods). This model assumes uniform protein concentrations within the ER and NE and focuses on protein movement through the junctions. The shape of the junctions were based on our previous measurements \cite{bragulat2024endoplasmic} (\textbf{Figure 2B}), and we used an averaged geometry for the mathematical model (\textbf{Figure 2A}, right panel).  To simplify the analysis, all junctions are assumed to share this average form. Because the junctions are approximated as long and narrow, we modeled them as one dimensional systems where protein movement follows a simplified diffusion equation. We used this model to simulate ER-to-NE diffusion of the two reporters and compared the results with our experimental data. All parameters required for the simulation were either measured experimentally or estimated based on data from previous studies. A summary of the parameter values used in the model is provided in \textbf{Table \ref{tab:table1}}. Because the length and angle of the junctions cannot be clearly determined from our EM images, we treated these features as variables in the model. We then compared the measured flux of ER-luminal reporters to the NE with model simulations at different junction lengths (5, 10, and 20 nm) and membrane angles (0°, 25°, and 50°).

\begin{table}[h!]
\centering
\begin{tabular}{|c|c|c|} 
 \hline
 \textbf{Parameter} & \textbf{Value} & \textbf{Description} \\
 \hline\hline
  $D_l$ & $0.52 \mu m^2/s$   & Diffusion coefficient of a large reporter   \\ 
  \hline
 $D_s$ & $3.3\mu m^2/s$
 & Diffusion coefficient of a small reporter \\ 
 \hline
 k & 40 & Number of junctions  \\
 \hline
 $V_{ER}$ & $200\mu m^3$ & Total volume of the ER  \\
 \hline
 $V_{NE}$ & $30\mu m^3$ & Volume of the NE  \\
 \hline
 $L$ & $5,\ 10,\ 20 \times 10^{-3}\ \mu\text{m}$ & Height of a junction \\
 \hline
 $R$ & $11 \times 10^{-3}\ \mu\text{m}$ & Average radius of a junction   \\
  \hline
 $r_s$ & $1.75 \, \times 10^{-3}\ \mu\text{m}$ & Average radius of the small reporter   \\
  \hline
 $r_l$ & $2.5 \times 10^{-3}\ \mu\text{m}$ & Average radius of the large reporter   \\
 \hline 
  $\alpha$ & $0^\circ,\, 25^\circ,\, 50^\circ $ & Opening angle of the truncated cone   \\
 \hline 
\end{tabular}
\caption{\textbf{Parameters values used in the mathematical model.} Apparent diffusion coefficients for the small and large protein reporters were obtained from FRAP measurements (\textbf{Figure 1B,C}). Volumes of the ER and the NE are based on data from volume electron microscopy studies \cite{Griffiths1984,Heinrich2021}. Junction number and radius were quantified using electron tomography, as described in a previous study \cite{bragulat2024endoplasmic}. Although two types of junctions were identified in that study (those with and without lumen), only lumen-containing junctions were included in the model, as junctions lacking lumen are unlikely to contribute to luminal ER-to-NE transport. A correction factor was applied to account for a 17\% shrinkage caused by the electron microscopy process.}
\label{tab:table1}
\end{table}

For the smaller reporter, the experimentally measured flux was slightly faster than the simulated values at a junction length of 10 nm and membrane angle of 25°, but remained closely aligned in magnitude (\textbf{Figure 2C,D}). Since these parameters fall within the observed structural range (\textbf{Figure 2B}), the model successfully captures the essential transport dynamics and approximates the correct order of magnitude. For the larger reporter, the measured flux deviates more noticeably from the simulated values, likely due to its elongated shape (\textbf{Figure 1A}). Unlike the globular smaller reporter, an elongated reporter may experience orientation-dependent diffusion, especially near membrane  surfaces or within confined spaces. Nonetheless, the observed flux of the larger reporter remained within the range of model predictions for junction lengths of 5–10 nm and angles of 25°–50° (\textbf{Figure 2E,F}), suggesting that its ER-to-NE flux can still be explained by simple diffusion.  
Note that the results shown in \textbf{Figure 2E,F} correspond to a radius of the larger reporter of $2.5 \times 10^{-3}\ \mu\text{m}$. \textit{A priori}, the estimated radius could range between $2.5 \times 10^{-3}$ and $6 \times 10^{-3}\ \mu\text{m}$ given the shape of the reporter; however, larger values still yield the correct order of magnitude.

Altogether, these results support the conclusion that simple diffusion is sufficient to explain the rapid transport of proteins from the ER to the NE, despite the narrow and limited geometry of the connecting junctions. We discuss potential biophysical and geometric factors contributing to the slightly faster-than-predicted flux in the Discussion section.

\begin{figure}[htbp]
    \centering
    \includegraphics[width=\textwidth]{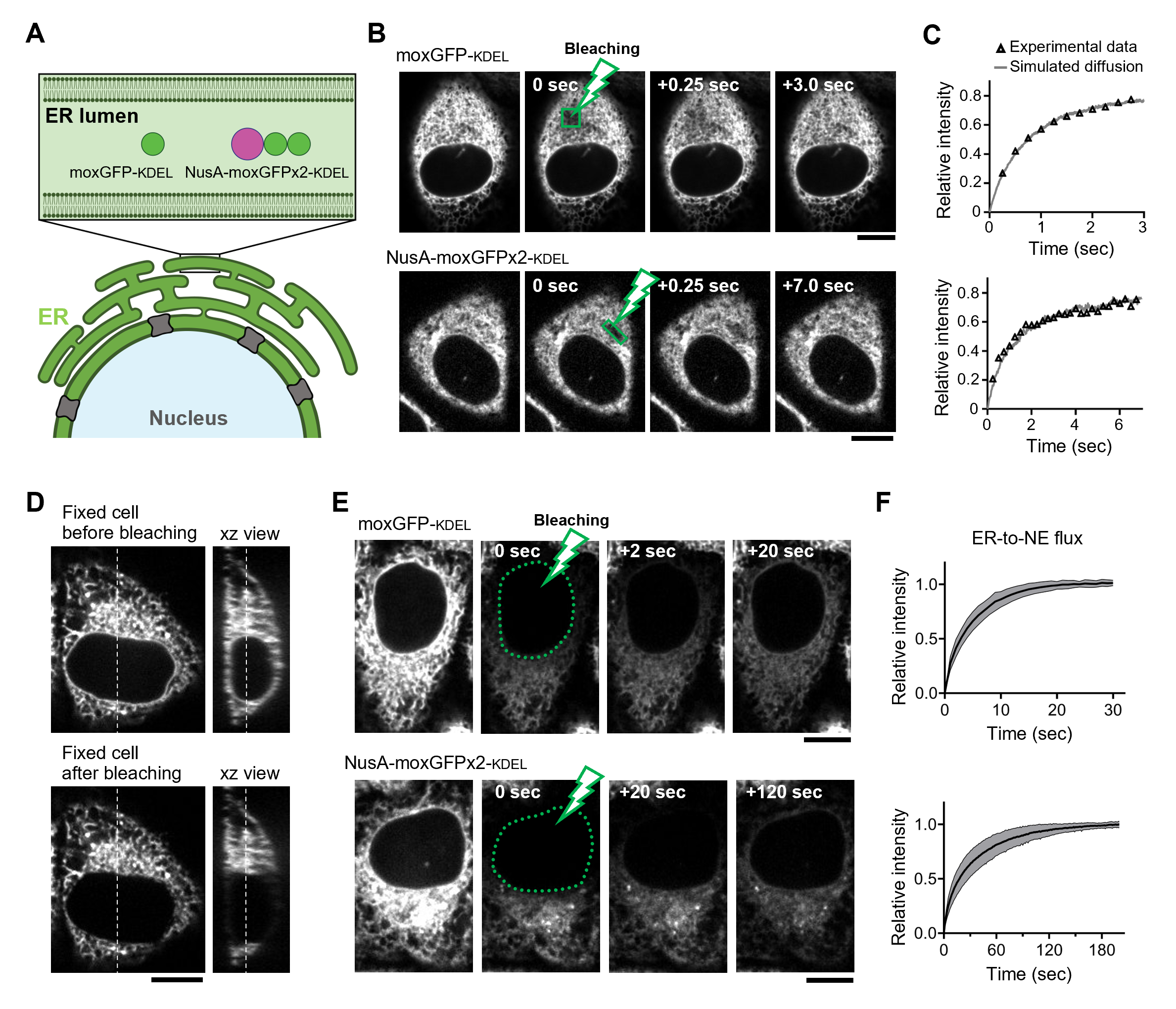}
    \caption{
        \textbf{Quantifying ER-to-NE transport of luminal proteins.}
        \textbf{(A)} Illustration of two reporters used to measure intraluminal protein transport between the ER and NE. 
        \textbf{(B)} Representative images of HeLa cells expressing the transport reporters: moxGFP-KDEL (top) and NusA-moxGFPx2-KDEL (bottom). 
        Diffusion within the ER was monitored by photobleaching a small region (indicated by the green line) and tracking fluorescence recovery over time. 
        \textbf{(C)} Fluorescence intensity in the bleached area was quantified for moxGFP-KDEL (top) and NusA-moxGFPx2-KDEL (bottom). 
        Experimental data (black triangles) and simulated diffusion curves (gray lines) for the cells shown in {(B)} are shown. 
        \textbf{(D)} Photobleaching in fixed cells to assess the degree of fluorescence loss across the entire NE, serving as a control for ER-to-NE transport measurements. 
        The left panels show a HeLa cell expressing moxGFP-KDEL before (top) and after (bottom) photobleaching the entire NE. 
        Orthogonal views of the indicated positions (white dashed lines) are shown on the right. 
        \textbf{(E)} ER-to-NE diffusion of two reporters was monitored following photobleaching of the entire NE, indicated by green dotted lines. 
        \textbf{(F)} Fluorescence intensity at the NE was quantified. Black and gray lines represent the mean and standard deviation (s.d.) of measurements from 40 cells across 4 independent experiments for moxGFP-KDEL (top) and 71 cells across 4 independent experiments for NusA-moxGFPx2-KDEL (bottom). 
        Scale bars: 10 $\mu$m.
    }
    \label{fig:fig1}
\end{figure}

\begin{figure}[htbp]
    \centering
    \includegraphics[width=\textwidth]{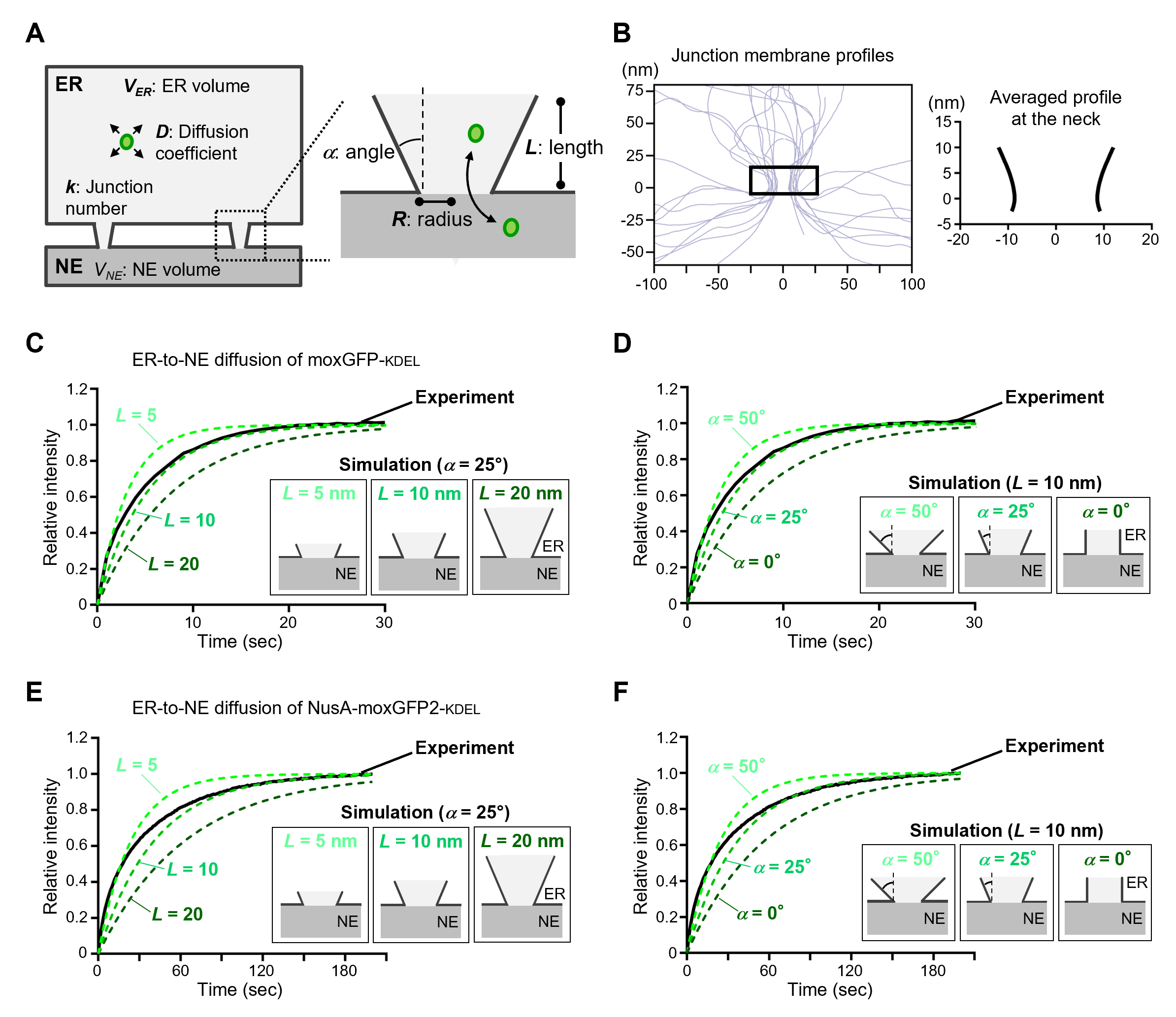}
    \caption{
        \textbf{The differential equation model with a geometric constraint predicted ER-to-NE flux that closely
matches the experimental data.}
\textbf{(A)} Schematic illustration of the parameters used in the mathematical model.
\textbf{(B)} Membrane profiles of the ER-NE junction from a previous study \cite{bragulat2024endoplasmic}. Although
two types of junctions were identified in that study (those with a visible lumen and those without), only lumen-
containing junctions were included in the model, as the non-luminal junctions are unlikely to support luminal
ER-to-NE transport. The averaged shape of the junction neck, highlighted by a bold rectangle in the left panel,
is shown in the right panel.
\textbf{(C,D)} Simulated ER-to-NE diffusion at different junction length \textbf{(C)} and membrane angles \textbf{(D)} (green dotted
lines) compared with experimental data from Figure 1E,F (black solid line) for moxGFP-KDEL.
\textbf{(E,F)} Simulated ER-to-NE diffusion at different junction length \textbf{(E)} and membrane angles \textbf{(F)} (green dotted
lines) compared with experimental data from Figure 1E,F (black solid line) for NusA-moxGFPx2-KDEL.
    }
    \label{fig:fig2}
\end{figure}

\newpage
\section{Discussion}
In this study, we combined live-cell imaging and computational modeling to investigate whether passive diffusion alone can account for luminal protein flux from the ER to NE. Although ER–NE junctions are both narrow and occur infrequently, our results show that simple diffusion is sufficient to explain the observed kinetics of protein accumulation in the NE following photobleaching. This finding addresses a central question in cell biology: how does luminal communication between the ER and NE occur efficiently despite limited architectural connectivity? While previous studies have described the continuity of the ER and NE \cite{Watson1955, Whaley1960, powell1990internuclear, Ellenberg1997, Pawar2021}, whether the continuity enables functionally sufficient transport had not been quantitatively tested. Our FRAP experiments reveal that proteins as large as 120 kDa can reach the NE from the ER within minutes, even when diffusion must occur through rare and narrow junctions. This suggests that the ER–NE system has evolved to permit efficient luminal equilibration without the need for active transport mechanisms.

The mathematical model we developed predicts the correct order of magnitude of the transport rate for both reporters across a physiologically realistic range of junction geometries. However, in both cases, the experimentally observed flux was slightly faster than predicted by the model (\textbf{Figure 2C–F}). These discrepancies are not unexpected, since the model makes several very rough approximations in order to allow
for explicit computations. It can be expected to capture the correct order of magnitude of the transport rate
rather than precise values. These discrepancies arise not only because of the approximations done in the model, but also from the variability in junction architecture. Indeed, the ER is highly dynamic \cite{nixon2016increased}, and we observed considerable variability in the ER membrane morphology adjacent to (approximately 10 nm above) the ER–NE junctions (\textbf{Figure 2B}). These structural fluctuations in junction length or angle could facilitate more efficient diffusion than static geometries allow.
 Our model assumes steady-state conditions and uniform junction geometry, which may not capture the spatial and temporal heterogeneity of live cells. Additionally, while the FRAP analysis provides robust average measurements of diffusion, it does not resolve single-molecule trajectories or identify subpopulations with distinct diffusion properties. Future studies combining live-cell-compatible super-resolution microscopy, such as MINFLUX \cite{gwosch2020minflux}, combined with single-particle tracking, could provide deeper insights into the dynamics of ER-to-NE transport. 
From the modeling point of view, the junction geometry enters the theoretical prediction through 
$$
\frac{A^*}{L} = \left(\int_0^L \frac{dz}{A(z)} \right)^{-1}\,.
$$
From EM representations of junctions, the points of connections with the ER are not
clearly defined, resulting in an ambiguity concerning the choice of the junction height $L$. We claim that this is not a very critical issue. Typically, the junction geometry is conical, and a model of the form $A(z) = (R+z\tan\alpha)^2\pi$ seems reasonable, with the radius $R$ at the connection to the NE and the opening angle $\alpha$ of the cone. This gives $A^* = R\pi(R+L\tan\alpha)$, and therefore $A^*/L$ tends to the limit $R\pi\tan\alpha$, as $L$ tends to infinity. This shows that the choice of $L$ does not have a strong influence, as long as it is chosen not too small.

Altogether, our results support a simple but robust mechanism of ER-to-NE protein transport driven by passive diffusion. These findings demonstrate that even minimal physical connectivity is sufficient for functionally effective luminal exchange and provide a quantitative basis for future studies exploring how ER–NE transport is modulated under various physiological conditions, developmental stages, or disease states. 


\section*{Acknowledgements}
This project was supported by Austrian Science Fund (FWF) (DOI: 10.55776/W1261) to S.M.A. and S.O., Austrian Science Fund (FWF) (DOI: 10.55776/F65) to S.M.A., Austrian Science Fund (FWF) (DOI: 10.55776/P28705) to C.M., European Research Council (ERC-2023-COG; 101124404 conNEctoER) to S.O., laboratory startup funding from the Medical University of Vienna to S.O., and the Vienna Science and Technology Fund (WWTF) (DOI: 10.47379/VRG17014) to S.M.A.. We acknowledge the BioOptics Light Microscopy Facility at the Max Perutz Labs (Thomas Peterbauer and Josef Gotzmann) for their expert support and training.
In addition, we thank the members of the labs of Shotaro Otsuka (especially Helena Bragulat-Teixidor, Clara-Anna Wagner, Kaike Ren, Tamara Völkerer and Michael Leichmann).
We used OpenAI’s ChatGPT (June 2025 version) to assist with writing refinement and language editing. The authors reviewed and edited all content to ensure accuracy and originality.

\newpage
\printbibliography

\newpage
\appendix
\section{Derivation of the mathematical model}
 \label{Scalingsection}

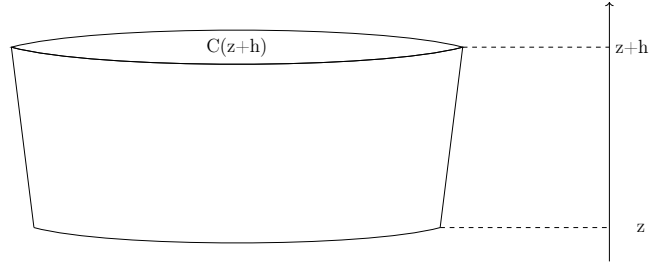
\begin{figure}[h!]
\centering
\scalebox{0.6}{
\begin{tikzpicture}
	\begin{pgfonlayer}{nodelayer}
		\node [style=none] (31) at (0, 18) {};
		\node [style=none] (32) at (0.5, 14) {};
		\node [style=none] (33) at (9.5, 14) {};
		\node [style=none] (34) at (10, 18) {};
		\node [style=none] (43) at (13.75, 18) {\large z+h};
		\node [style=none] (45) at (13.95, 14) {\large z};
        \node [style=none] (46) at (5, 18) {\large C(z+h)};
		\node [style=none] (47) at (9.75, 15.75) {};
		\node [style=none] (48) at (11.25, 15.5) {};
		\node [style=none] (50) at (13.25, 19) {};
		\node [style=none] (51) at (13.25, 13.25) {};
		\node [style=none] (53) at (13.25, 18) {};
		\node [style=none] (54) at (13.25, 14) {};
	\end{pgfonlayer}
	\begin{pgfonlayer}{edgelayer}
		\draw [style=none] (34.center)
			 to [bend right=15, looseness=0.50] (31.center)
			 to [bend right=15, looseness=0.50] cycle;
		\draw [style=Arrow] (51.center) to (50.center);
		\draw [style=Dashed] (34.center) to (53.center);
		\draw [style=Dashed] (33.center) to (54.center);
		\draw [style=none] (31.center) to (32.center);
		\draw [style=none] (33.center) to (34.center);
		\draw [style=none, bend right=15, looseness=0.50] (31.center) to (34.center);
		\draw [bend right=15, looseness=0.50] (32.center) to (33.center);
	\end{pgfonlayer}
\end{tikzpicture}}
\caption{\small Cross-sectional slice of a junction} \label{fig:Domain3}
\end{figure}

The derivation of \eqref{rhoJ} starts with the three-dimensional diffusion equation
$$
   \frac{\partial\rho_J}{\partial t} = D\Delta\rho_J = D\left( \frac{\partial^2\rho_J}{\partial x^2} + \frac{\partial^2\rho_J}{\partial y^2} +\frac{\partial^2\rho_J}{\partial z^2}\right) \,,
$$
holding within the junction, where $z$ is the longitudinal position variable, as above, and cross sections are parallel to the $(x,y)$-plane. The dimension reduction asymptotics is a classical procedure for problems with extreme aspect ratios. Here we assume that the average radius $R^*$ of the junction is small compared to its height $L$, i.e., that the aspect ratio
$\eps = R^*/L$ is small. In the diffusion equation the position variables and time are non-dimensionalized by
$$
   (x,y) \to R^*(x,y) \,,\qquad z \to Lz \,,\qquad t \to \frac{L^2}{D} t \,,
$$
leading to the scaled version
\be\label{diff-scaled}
      \eps^2\frac{\partial\rho_J}{\partial t} = \frac{\partial^2\rho_J}{\partial x^2} + \frac{\partial^2\rho_J}{\partial y^2} + \eps^2\frac{\partial^2\rho_J}{\partial z^2} \,.
\ee
For a thin cross-sectional slice lying between $z$ and $z+h$ (see Figure \ref{fig:Domain3}), changes of mass are due to the flow through the bottom and top cross sections, denoted by $C(z)$ and, respectively, $C(z+h)$.

Mathematically, this is the consequence of applying zero flux boundary conditions along the walls of the junction, integrating \eqref{diff-scaled} (divided by $\eps$) over the slice, and using the divergence theorem:
\begin{eqnarray}
    &&\frac{d}{dt} \int_z^{z+h} \int_{C(\zeta)} \rho_J(x,y,\zeta,t) d(x,y)d\zeta \nonumber\\
    && = \int_{C(z+h)} \frac{\partial\rho_J}{\partial z}(x,y,z+h,t) d(x,y) - \int_{C(z)} \frac{\partial\rho_J}{\partial z}(x,y,z,t) d(x,y) \,.\label{diff-int}
\end{eqnarray}
The limit of \eqref{diff-scaled} as $\eps\to 0$ is the two-dimensional Laplace (stationary diffusion) equation
$$
   \frac{\partial^2\rho_J}{\partial x^2} + \frac{\partial^2\rho_J}{\partial y^2} = 0 \,,
$$
which, together with no-flux boundary conditions through the junction walls, implies that 
the density is independent of $(x,y)$, thus $\rho_J = \rho_J(z,t)$. Therefore, passing to the limit $\eps\to 0$ in \eqref{diff-int} gives
$$
    \int_z^{z+h} A(\zeta)\frac{\partial\rho_J}{\partial t}(\zeta,t)d\zeta 
   = A(z+h)\rho_J(z+h,t) - A(z)\rho_J(z,t) \,.
$$
The final steps in the derivation of the nondimensionalized version of \eqref{rhoJ} are division by $h$ and passage to the limit $h\to 0$.

The second goal is to derive \eqref{rhoNE}--\eqref{rhoERx} as an approximation for \eqref{rhoJ}--\eqref{BC}. Again the first step is a non-dimensionalization (of \eqref{rhoJ}--\eqref{rhoNE}). However, now we concentrate on the time scale relevant for the dynamics in the NE:
$$
  z\to Lz \,,\qquad t \to \frac{V_{NE} L}{kDA^*} t \,,\qquad A\to A^* A \,.
$$
The scaled version of \eqref{rhoJ}--\eqref{rhoNE} then reads
\begin{eqnarray*}
   \delta_1 A(z)\frac{\partial \rho_J}{\partial t}(z,t) &=& \frac{\partial}{\partial z}\left( A(z) \frac{\partial\rho_J}{\partial z}(z,t)\right) \,,\\
   \frac{d\rho_{ER}}{dt}(t) &=& -\delta_2 A(1) \frac{\partial\rho_J}{\partial z}(1,t) \,,\\
   \frac{d\rho_{NE}}{dt}(t) &=& A(0) \frac{\partial\rho_J}{\partial z}(0,t) \,,
\end{eqnarray*}
with the dimensionless parameters
$$
  \delta_1 = \frac{kLA^*}{V_{NE}} \,,\qquad \delta_2 = \frac{V_{NE}}{V_{ER}} \,,
$$
the ratios of the total junction volume to the NE volume and, respectively, the NE volume to the total ER volume. As mentioned in Section \ref{sec:mat-meth}, both these ratios are small,
justifying the approximation $\delta_1=\delta_2=0$, leading to the scaled version of \eqref{rhoNE}--\eqref{rhoERx}.

\end{document}